# Band Gap of Strained Graphene Nanoribbons.

Yang Lu and Jing Guo
Department of Electrical and Computer Engineering, University of Florida, Gainesville, FL, 32611

ABSTRACT

The band structures of strained graphene nanoribbons (GNRs) are examined by a tight binding Hamiltonian that is directly related to the type and strength of strains. Compared to the two-dimensional graphene whose band gap remains close to zero even if a large strain is applied, the band gap of graphene nanoribbon (GNR) is sensitive to both uniaxial and shears strains. The effect of strain on the electronic structure of a GNR strongly depends on its edge shape and structural indices. For an armchair GNR, uniaxial weak strain changes the band gap in a linear fashion, and for a large strain, it results in periodic oscillation of the band gap. On the other hand, shear strain always tend to reduce the band gap. For a zigzag GNR, the effect of strain is to change the spin polarization at the edges of GNR, thereby modulate the band gap. A simple analytical model is proposed to interpret the band gap responds to strain in armchair GNR, which agrees with the numerical results.

## I. Introduction

Strain has been extensively used in silicon electronics industry for boosting the device performance and has played an important role since the 90nm technology node [1].Compared to silicon, graphene is an atomically thin two-dimensional (2D) material and therefore is structurally more amenable to external modifications including strain. The graphene material can sustain a much larger strain compared to the silicon material. The effect of strain on 2D graphene has been studied both experimentally and theoretically, including the effects of uniform [2-5] and local strains[6]on the electronic structure, as well as the possibility to achieve the quantum Hall states in the absence of the external magnetic field[7].The 2D graphene does not have a band gap, and the band gap remains close to zero even if a strain as large as 20% is applied. A band gap can be created by patterning the 2D graphene to a nanometer-wide graphene nanoribbon (GNR), this is predicted theoretically [8-10] and realized experimentally [11-13]. GNRs present interesting transport properties where, for example, disorder such as imperfect edges, can play an important role [14]. Moreover, strain could be useful to further tailor the electronic properties of GNRs. Based on density functional theory (DFT), the effect of uniaxial strain on the electronic properties of GNRs have been studied before [15][16]. Both work revealed the potential of uniaxial strain in tuning GNR's electronic peroperties. The underlying physics of strain effects on the band gap of GNRs, however, is buried in DFT simulations and is not fully understood.

In this work, a systematic study on effect of both uniaxial and shear strain on the bandgap of GNRs is performed by a tight binding Hamiltonian that is directly related to the strength and type of strain. An analytical model is developed to describe the dependence of bandgap on strain in AGNRs. The work provides explicit relations between the bandgap and strain in GNRs, which enables simple and detailed physical understandings. It is observed that the band gap of a GNR is much more sensitive to strain than 2D graphene and strongly depends on its edge shape and structural indices. For zigzag GNRs (ZGNRs), uniaxial and shear strain modulate the spin density at the GNR edges thereby alter the band gap. For armchair GNRs (AGNRs), uniaxial strain and shear strain result in qualitatively different dependence of the band gap on strain. The effect of strain on the band gap is qualitatively different for AGNRs with different structural indices. The effects of edge bond relaxation [9] and third nearest neighbor coupling [17] modify the quantitative dependence of the band gap on strain.

## II Approach

The bandstructures of the modeled GNRs are calculated by using a tight binding model, whose binding parameters have been parameterized by *ab initio* calculations in previous studies of GNR band structures in the absence of strain [9, 17]. For AGNRs, modeling the edge bond relaxation and the 3rd nearest neighbor coupling are necessary for treating the edge effects and describing all semiconducting band structures[13]as predicted by the *ab initio* calculations[9,10,17]. For ZGNRs, inclusion of a Hubbard term in the Hamiltonian is needed to describe the edge spin polarization and opening of the band gap[19]. The binding parameters in the presence of strain are modified according to the Harrison binding parameter relation. This approach has been used and validated before in the study of strain effects on carbon nanotubes[20].

As shown in Fig. 1(a), the unstrained bond vectors for an AGNR are given by,

$$\begin{aligned}\vec{r}_1 &= a_0 \hat{x} \\ \vec{r}_2 &= \tfrac{1}{2} a_0 \hat{x} + \tfrac{\sqrt{3}}{2} a_0 \hat{y}, \\ \vec{r}_3 &= \tfrac{1}{2} a_0 \hat{x} - \tfrac{\sqrt{3}}{2} a_0 \hat{y}\end{aligned} \qquad (1)$$

where we set $\hat{x}$ as the transport direction of GNR. The application of a uniaxial or shear strain causes the following changes,

$$\begin{aligned}r_{ix} &\to (1+\sigma) r_{ix} \\ r_{iy} &\to (1+\nu\sigma) r_{iy}\end{aligned} \quad \text{(Uniaxial strain)} \qquad (2)$$

$$r_{ix} \to r_{ix} + \gamma r_{iy}, \quad \text{(Shear strain)}$$

where $i=1,2,3$ and $r_{ix}$, $r_{iy}$ are the x and y component of $\vec{r}_i$. $\sigma$ represent the uniaxial strain along x direction, $\nu \approx 0.165$ is the Poisson ratio[21], and $\gamma$ is the shear strain. Here we focus on the simple case of uniform strain, while in practice the deformation of graphene may include long range and regular ripples patterns. A tight-binding Hamiltonian as parameterized by Gunlycke and White[22], which includes the treatment of the edge bond relaxation and the 3rd nearest neighbor coupling, is used to compute the band structure of the AGNR. In the presence of strain, each binding parameter is scaled by a dimensionless factor $\xi = (\frac{r_0}{r})^2$, where $r_0$ the unstrained bond length, and $r$ is the bond length in the presence of strain.

The bond lengths of the zigzag GNR, as shown in Fig. 1(b), are modified in a similar manner as the AGNR in the presence of strain. Due to the existence of localized edge states in zigzag GNR, the spin polarized interaction should be included in the Hamiltonian of the system, which can be generally described as[23],

$$H = \sum_{\langle i,j\rangle,\sigma} t_{ij} c_{i\sigma}^{+} c_{j\sigma} + U\sum_{i,\sigma}(\langle n_{i,-\sigma}\rangle - \frac{1}{2})n_{i,\sigma}, \qquad (3)$$

where $c_{i\sigma}^{+}, c_{j\sigma}$ and $n_{i\sigma}$ are creation, annihilation, and number operators, respectively, for an electron of spin $\sigma$ in the $\pi$-orbital centered on the *ith* C atom in the ribbon. $\langle i, j\rangle$ denotes the set of all nearest neighbors, $t_{ij}$ is the corresponding nearest neighbor hopping parameter and $U$ describe the strength of the spin dependent field. $\langle n_{i,\sigma}\rangle$ is the average electron density with spin $\sigma$ at the location of *ith* C atom, and can be calculated self-consistently from equilibrium carrier statistics. The Hamiltonian described by Equation(6) is in fact equivalent to Hatree-Fock approximation applied to Hubbard model[24], This is first studied by Fujita et al in his paper about edge states in zigzag GNRs[25].

## III. Results and Discussion

### 3. 1 Armchair GNR

We first consider the case of uniaxial strain. The band structure $E_n(k)$ is calculated and the band gap is obtained by finding the minimum of $2|E_n(k)|$ for all the band index $n$ and wave vector $k$. For the purpose of comparison, we first neglect the effect of edge bond relaxation and third nearest neighbor coupling. In this simple case, the band structure of a strained AGNR with an index of $n$ is similar to that of a strained zigzag SWNT with an index of $n+1$[17], except the lack of valley degeneracy. As shown in Fig.2 (a), the band gap scales linearly with the strain strength in certain range, and repeats itself periodically as the strain strength further increases. The effect of strain on band gap is significant and qualitatively different between the cases of *n+1=3q*, *3q+1* and *3q+2* AGNR. For *n=23* (*3q*) case, small tensile strain increase the band gap, with only 5% uniaxial strain, it opens a band gap up to about 0.4eV. On the other hand, small tensile strain increase the band gap of *n=24* AGNR (*3q+1*), and decrease the band gap of *n=25* AGNR (*3q+2*).

In the presence of edge bond relaxation and the 3rd nearest neighbor coupling, the band gap is non-zero in the absence of strain for any AGNR. Figure 2(b) plots the band gap of AGNR under uniaxial strain when the effect of edge bond relaxation and 3rd nearest neighbor coupling are included. The qualitative features of band gap to strain relation do not change, but the quantitative value of the band gap is perturbed. Furthermore, the maximum achievable band gap in the presence of compressive strain is smaller than that in the presence of tensile strain.

In order to obtain a simple relation between strain and the band gap of AGNR, we can calculate the lowest order contribution of strain to band structures, the Eigenenergies of an AGNR at *k=0* can be written as (see appendix),

$$E(p) = t_0(1+\alpha+2(1+\beta)\cos\frac{p\pi}{n+1}) + t_{3n}(1+\alpha+2(1+\beta)\cos\frac{2p\pi}{n+1})$$
$$+ \frac{4(t_{3n}(1+\beta)+\Delta t_e)}{n+1}\sin^2\frac{p\pi}{n+1}$$  (4)

where
$$\alpha = -2\sigma + 3\sigma^2$$
$$\beta = -\frac{(1-3\nu)\sigma}{2} + \frac{3}{4}\gamma^2 + \frac{1}{4}(1-3\nu)^2\sigma^2.$$  (5)

Here, $p$ is band index running from *1* to *n*, $t_0$ is the nearest neighbor hopping integral in the absence of strain, $\sigma$ and $\gamma$ are the strength of uniaxial strain and shear strain, respectively, $\nu$ is the Poisson ratio, $t_{3n}$ is the 3rd nearest neighbor coupling strength and $\Delta t_e$ is the correction due to edge bond relaxation. In equation (4), the first term corresponds to band energy neglecting edge bond relaxation and third nearest neighbor coupling, the second and third term account for these two effects. $\alpha, \beta$ are the corrections due to strain. Equation (4) gives *n* of the *2n* Eigenenergies of the system. The other *n* eigenenergies, due to symmetry, are just the opposites of Equation (4). Thus, the band gap can be calculated as,

$$E_g = \min_{P=1,2..n} |2E(p)|.$$  (6)

From Eq. (4), it is observed that there is no first order contribution to the band gap from shear strain. Also, because $t_{3n}$ and $\Delta t_e$ is relatively small compared to $t_0$, we then preserve only the first term of equation (4), which is the dominating factor for the qualitative features of the band gap dependence on strain,

$$E_g \approx 2|t_0| \min_{P=1,2..N} |(1+\alpha + 2(1+\beta)\cos\frac{p\pi}{n+1})| . \tag{7}$$

To the first order of uniaxial strain, Equation (7) can be further approximated as,

$$E_g \approx \min_{p=1,2...N} (-2\sqrt{3}\pi t_0) |\frac{p}{n+1} - p_0|, \tag{8}$$

where $p_0 = \frac{2}{3} - \frac{\sqrt{3}}{2\pi}(1+v)\sigma$.

Equation (8) implies that the band gap is proportional to the shortest distance between $p_0$ and $\frac{p}{n+1}$, the quantization grids in the width direction of the GNR. This is depicted in Fig.3. It's observed that when $n+1=3q$, the shortest distance of $p_0$ to the grids is 0, and for n+1=3q+1 or 3q+2, the shortest distance *of* $p_0$ to the grids is $\frac{1}{3(n+1)}$. This explains the features of Fig.2, as for all three cases, small tensile strain shift the position of $p_0$ to the negative direction, which result in the minimum distance *of* $p_0$ to grids is increased for *n+1=3q or 3q+1,*and decreased for *n+1=3q+2.* The behavior in compressive strain can be explained similarly. Fig.3. also provides an explanation for the qualitatively periodic oscillation of the band gap as the strain increases. Further shifting the position of $p_0$ results in a periodical repetition of that minimum distance, which gives the periodical pattern of band gap versus strain relations. Furthermore, The maximum achievable band gap is proportional to half the grid space, $\frac{1}{2(n+1)}$. For *n=3q or 3q+1* AGNR, this corresponds to about 50% increase of the band gap compared to the unstrained case. The explanation of band gap oscillation under strain is similar to previous studies of strain effects on nanotubes, which attribute the change of band gap to the shifting of Fermi point under strain.[18,20]

We also plot the dependence of band gap on both strain and ribbon width, as shown in Fig.4. The simulated range of uniaxial strain strength is from -15% to 15%, and the width from *2* to *10* nm. The periodic oscillations of the band gap as a function of the uniaxial strain strength and the qualitative difference between *3q*, *3q+1* and *3q+2* groups are observed for the whole range of the simulated parameters. In general, increasing the width of ribbon reduces the maximum achievable band gap, due to weaker confinement in the width direction.

The effect of shear strain on band gap of AGNR is qualitatively different from that of uniaxial strain, as shown in Fig.5. As equation (4) and (5) indicates, there is no first order contribution from shear strain to band gap, so the dependence of the band gap on shear strain is due to the second and higher order perturbation effects. In the presence of edge bond relaxation and the 3rd nearest neighbor coupling, shear strain always reduce the band gap regardless of the structural indices of the AGNR.

**3.2 Zigzag GNR.**

The band gap of ZGNR originates from totally different mechanism as compared to AGNR. As indicated by equation (3). The spin interaction is included and the band separation at zone boundary can be approximated as [19],

$$\Delta E = U(n_{1,\uparrow} - n_{1,\downarrow}). \tag{9}$$

The actual band gap is proportional but smaller than $\Delta E$. In Fig.6, we plot the band structure of $n=16$ ZGNR. The blue solid line is the unstrained band structure while the red dashed line corresponds to band structure under 15% uniaxial strains. Obviously, this tensile strain opens up the band gap. Then we calculate the dependence of ZGNR's band gap on uniaxial strain, as shown in Fig.7 (a). In contrast to AGNRs, the band gap of a ZGNR increases as tensile strain is applied and decreases as compressive strain is applied regardless of its structural index. The normalized band gap $\frac{E_g}{E_{g0}}$ versus strain is shown in the inset of Fig.7 (a), which is approximately the same for ZGNR with different widths. We fit the curve and get the empirical relation,

$$\frac{E_g}{E_{g0}} = \sigma^2 + 1.6\sigma + 1, \tag{10}$$

where $E_{g0}$ is the unstrained band gap. To explain the effect of strain on band gap, we calculated the edge spin polarization for various ZGNR under uniaxial strain. As shown in Fig.7(b), positive(negative) uniaxial strain always tend to increase(decrease) the edge spin polarization. As indicate by equation (14), stronger spin polarization will induce larger band separation, which is roughly proportional to the band gap. This justifies the monotonous feature of band gap versus strain in Fig.7.

The reason why tensile strain will increase edge spin polarization can be conceptually captured by the analytical model proposed by Fujita[25], For edge states, the corresponding charge density is proportional to $[2\cos(k/2)]^{2m}$ at each non-nodal site of the $mth$ zigzag chain from the edge. So $2\cos(k/2)$ represent the "damping length" of the edge states. If a strain is applied, due to the distortion of bond vector and bond parameter, this damping factor should be modified as $2\frac{t_2}{t_1}\cos(k/2)$, where $t_1, t_2$ are bond parameters relate to bond vectors $\vec{r}_1, \vec{r}_2$ in Fig.1(b). For tensile strain $(\sigma > 0)$, $|t_2/t_1| < 1$, making the damping of edge states much quicker, which result in more localized edge states. Due to electron-electron interaction, this will increase the spin polarization at edge sites, thereby increasing the band gap of the system. A Similar argument applies in the case of compressive strain $(\sigma < 0)$.

We also calculate the effect of shear strain on the band gap of ZGNR, as shown in Fig.8 (a). Compared to the case of uniaxial strain, the change of band gap is relatively small, and shear strain always tend to reduce the band gap. These features could also be explained by Fujita's model [25]. We find that, under shear strain, the damping factor should be modified as $2\cos(\frac{k}{2})\sqrt{1+\frac{3}{4}\gamma^2\tan^2(\frac{k}{2})}$, where $\gamma$ is the shear strain strength. Because $\sqrt{1+\frac{3}{4}\gamma^2\tan^2(\frac{k}{2})} \geq 1$, under shear stain, the damping of edge states is slower than without strain. So the edge states are less localized, thus decreasing the spin polarization at edge sites, and therefore reducing the band gap. This analysis is confirmed by Fig.8 (b), in which shear strain always reduce the spin polarization at edges.

## IV conclusions

In conclusion, we explored the effect of strain on the band gap of GNR. Two types of strain, (uniaxial, shear strain) and two types of GNRs (AGNR and ZGNR) are studied. The effect of strain is modeled as a modification to the tight-binding nearest neighbor hopping integral. It is found that, for AGNR, uniaxial strain linearly shifts the band gap, which periodically repeats itself as strain strength is further increased. Shear strain has no obvious contribution in opening up the band gap. In all cases, it tends to reduce the band gap. We explained these observations by proposing a perturbation model and it well reproduced the results of numerical calculations. For ZGNR, we find that strain changes the spin polarization at edge sites of nanoribbon, thus further affecting the band gap. Tensile strain increases the band gap while compressive and shear strain reduce the band gap. These results indicate that, the band gap of GNR is sensitive to strength of strain. By applying moderate strain strength, the electronic properties of GNR can be readily engineered.

## Acknowledgement

This work was supported by ONR and NSF.

**Appendix**: the derivation of equation (4)

For AGNR, in the Tight Bind model, due to the symmetry of Hamiltonian, the band gap always occurs at $k=0$. At $k=0$, the Tight Bind Hamiltonian reduce to a 2 leg ladder lattice system [9], as shown in Fig.A.1

$$H_0 = \begin{pmatrix} T & t_0 I \\ t_0 I & T \end{pmatrix}, \tag{A.1}$$

where $T = \begin{pmatrix} 0 & t_0 & & & \\ t_0 & 0 & t_0 & & \\ & t_0 & ... & t_0 & \\ & & t_0 & 0 & \end{pmatrix}$,

corresponds to the left (right) leg of the ladder.

The Eigenstates and Eigenenergies of $H_0$,

$$H_0 |E_p\rangle = E_p |E_p\rangle \quad , \quad E_p = t_0(1+2\cos\frac{p\pi}{n+1}), \quad p=1,2..n \;, \tag{A.2}$$

where $|E\rangle = \begin{pmatrix} \psi_p \\ \psi_p \end{pmatrix} \quad , \quad \psi_p = \begin{pmatrix} \sin\frac{p\pi}{n+1} \\ \sin\frac{2p\pi}{n+1} \\ ... \\ \sin\frac{np\pi}{n+1} \end{pmatrix}.$

Equation (A.2) describes $n$ of the $2n$ Eigenstates we would like to discuss. The other $n$ Eigenstates, due to symmetry, just have the opposite eigenenergies, $-E_p$, $p=1,2...n$.

The perturbed Hamiltonian and energy due to strain,

$$H' = \begin{pmatrix} T_L & dt_1 I \\ dt_1 I & T_R \end{pmatrix}, \quad T_L = \begin{pmatrix} 0 & dt_3 & & \\ dt_3 & 0 & dt_2 & \\ & dt_2 & ... & dt_3 \\ & & dt_3 & 0 \end{pmatrix}, \quad T_R = \begin{pmatrix} 0 & dt_2 & & \\ dt_2 & 0 & dt_3 & \\ & dt_3 & ... & dt_2 \\ & & dt_2 & 0 \end{pmatrix},$$

$$\Delta E_p = \frac{\langle E_p | H' | E_p \rangle}{\langle E_p | E_p \rangle} = \frac{\begin{pmatrix} \psi_p^+ & \psi_p^+ \end{pmatrix} \begin{pmatrix} T_L & dt_1 I \\ dt_1 I & T_R \end{pmatrix} \begin{pmatrix} \psi_p \\ \psi_p \end{pmatrix}}{2\psi_p^+ \psi_p} = \frac{\psi_p^+ (T_L + T_R) \psi_p + 2 dt_1 \psi_p^+ \psi_p}{2\psi_p^+ \psi_p}$$

$$= dt_1 + \frac{\psi_p^+ (T_L + T_R) \psi_p}{2\psi_p^+ \psi_p} = dt_1 + 2 \frac{(dt_2 + dt_3)}{(N+1)} \sum_{k=1}^{N-1} \sin \frac{kp\pi}{n+1} \sin \frac{(k+1)p\pi}{n+1} \qquad (A.3)$$

$$= dt_1 + (dt_2 + dt_3) \cos \frac{p\pi}{n+1},$$

Using equation (2) and (3) we get the expression for the perturbed bonding parameter,

$$dt_1 = t_0(-2\sigma + 3\sigma^2)$$

$$dt_2 = t_0(+\frac{\sqrt{3}}{2}\gamma - \frac{(1-3\nu)\sigma}{2} + \frac{3}{4}\gamma^2 + \frac{1}{4}(1-3\nu)^2 \sigma^2 - \frac{\sqrt{3}}{2}(1-3\nu)\gamma\sigma) \quad . \qquad (A.4)$$

$$dt_3 = t_0(-\frac{\sqrt{3}}{2}\gamma - \frac{(1-3\nu)\sigma}{2} + \frac{3}{4}\gamma^2 + \frac{1}{4}(1-3\nu)^2 \sigma^2 + \frac{\sqrt{3}}{2}(1-3\nu)\gamma\sigma)$$

Substitute (A.4) into (A.3), we have,

$$\Delta E_p = t_0 [(-2\sigma + 3\sigma^2) + ((3\nu - 1)\sigma + \frac{1}{2}(1-3\nu)^2 \sigma^2 + \frac{3}{2}\gamma^2) \cos \frac{p\pi}{n+1}] \quad . \qquad (A.5)$$

To include the effect of edge distortion and 3$^{rd}$ nearest neighbor coupling, we use similar method as above, the perturbed Hamiltonian due to 3rd nearest neighbor coupling and edge bond relaxation,

$$H_{3n} = \begin{pmatrix} 0 & T_{3n} \\ T'_{3n} & 0 \end{pmatrix} \quad , \quad T_{3n} = \begin{pmatrix} t_{3n}^{(1)} & 0 & t_{3n}^{(2)} & & & & & \\ 0 & t_{3n}^{(1)} & 0 & t_{3n}^{(3)} & & & & \\ t_{3n}^{(3)} & 0 & t_{3n}^{(1)} & 0 & t_{3n}^{(2)} & & & \\ & t_{3n}^{(2)} & 0 & t_{3n}^{(1)} & 0 & \ddots & & \\ & & t_{3n}^{(3)} & 0 & \ddots & \ddots & & t_{3n}^{(3)} \\ & & & \ddots & \ddots & t_{3n}^{(1)} & 0 \\ & & & & t_{3n}^{(2)} & 0 & t_{3n}^{(1)} \end{pmatrix},$$

$$H_{edge} = \begin{pmatrix} 0 & T_e \\ T'_e & 0 \end{pmatrix} \quad , \quad T_e = \begin{pmatrix} \Delta t_e & & & & \\ & 0 & & & \\ & & \ddots & & \\ & & & 0 & \\ & & & & \Delta t_e \end{pmatrix}$$

where $\Delta t_e$ is edge bond correction and $t_{3n}^{(1)}, t_{3n}^{(2)}, t_{3n}^{(3)}$ are $3^{rd}$ nearest neighbor coupling parameters under strain, which is given by expressions similar to (A.4),

$$\begin{aligned} t_{3n}^{(1)} &= t_{3n}(1 - 2\sigma + 3\sigma^2) \\ t_{3n}^{(2)} &= t_{3n}(1 + \frac{\sqrt{3}}{2}\gamma - \frac{(1-3\nu)\sigma}{2} + \frac{3}{4}\gamma^2 + \frac{1}{4}(1-3\nu)^2\sigma^2 - \frac{\sqrt{3}}{2}(1-3\nu)\gamma\sigma) \quad , \\ t_{3n}^{(3)} &= t_{3n}(1 - \frac{\sqrt{3}}{2}\gamma - \frac{(1-3\nu)\sigma}{2} + \frac{3}{4}\gamma^2 + \frac{1}{4}(1-3\nu)^2\sigma^2 + \frac{\sqrt{3}}{2}(1-3\nu)\gamma\sigma) \end{aligned} \quad (A.5)$$

where $t_{3n}$ is the unstrained 3rd nearest neighbor coupling parameter.

The perturbed energy,

$$\begin{aligned} \Delta E_{p3n} &= \frac{\langle E_p | H_{3n} | E_p \rangle}{\langle E_p | E_p \rangle} = \frac{\psi_p^+ (T_{3n} + T'_{3n}) \psi_p}{n+1} \\ &= t_{3n}^{(1)} + (t_{3n}^{(2)} + t_{3n}^{(3)}) \cos\frac{2p\pi}{n+1} + \frac{2(t_{3n}^{(2)} + t_{3n}^{(3)})}{n+1} \sin^2 \frac{p\pi}{N+1} \\ &= t_{3n}(1 - 2\sigma + 3\sigma^2) + 2t_{3n}(1 - \frac{(1-3\nu)\sigma}{2} + \frac{3}{4}\gamma^2 + \frac{1}{4}(1-3\nu)^2\sigma^2)\cos\frac{2p\pi}{n+1} \\ &\quad + \frac{4t_{3n}}{n+1}(1 - \frac{(1-3\nu)\sigma}{2} + \frac{3}{4}\gamma^2 + \frac{1}{4}(1-3\nu)^2\sigma^2)\sin^2\frac{p\pi}{n+1} \end{aligned} \quad (A.6)$$

$$\Delta E_{pe} = \frac{\langle E_p | H_{edge} | E_p \rangle}{\langle E_p | E_p \rangle} = \frac{2\psi_p^+ T_e \psi_p}{n+1} = \frac{4\Delta t_e}{n+1} \sin^2 \frac{p\pi}{n+1}, \tag{A.7}$$

Add (A.2), (A.5), (A.6) and (A.7) we get,

$$E(p) = t_0(1+\alpha+2(1+\beta)\cos\frac{p\pi}{n+1}) + t_{3n}(1+\alpha+2(1+\beta)\cos\frac{2p\pi}{n+1})$$
$$+ \frac{4(t_{3n}(1+\beta)+\Delta t_e)}{n+1} \sin^2 \frac{p\pi}{n+1} \tag{A.8}$$

where
$$\alpha = -2\sigma + 3\sigma^2$$
$$\beta = -\frac{(1-3\nu)\sigma}{2} + \frac{3}{4}\gamma^2 + \frac{1}{4}(1-3\nu)^2\sigma^2 \tag{A.9}$$

To see the effectiveness of this analytical approximation, we compared the results calculated by the TB model numerically in section 2 with that of equation (9) to (11), as figure A.2 shows.

# Figure Captions

Fig.1. (a) The unit cell of AGNR (b) unit cell of ZGNR. In each figure, **r1, r2, r3** are the bond vectors, the transport direction of the ribbon is set as x direction.

Fig.2. Band Gap versus uniaxial strain for AGNR, *n=23, 24, 25* corresponds to *n+1=3q, 3q+1* and *3q+2* AGNR respectively. (a) is the simple case neglecting edge bond relaxation and 3rd nearest neighbor coupling effect and (b) includes these two effects. In both figure, the band gap versus strain curve show similar periodic pattern. Locally, Band gap changes linearly when increase (decrease) strain strength.

Fig.3. Visualization of the position of $p_0$ in equation (13) in relative to the quantization grids, $\frac{p}{n+1}$, where *p=1 to n*. the distance of $p_0$ to nearest grid points is different for AGNR with different indices. For *n+1=3q*, it is *0*; for *n+1=3q+1* or *3q+2*, it is one third of the grid space, $\frac{1}{3(n+1)}$. As indicated by the red arrow, tensile strain shift $p_0$ to negative direction.

Fig.4. Band Gap versus uniaxial strain for different width of AGNR, with the effect of edge bond relaxation and third nearest neighbor coupling included. The width of AGNR is from 2nm to 10 nm. Generally, the band gaps still have a periodic dependence on strain strength, and are roughly inverse proportional to the width of the ribbons.

Fig.5. Band Gap versus shear strain for AGNR with edge bond relaxation and third nearest neighbor effect included. In this case, Shear strain always tends to reduce the band gap.

Fig.6. Band structure of *n=10* ZGNR, the solid blue line is the case without strain, dashed red line is the case with 15% uniaxial strain. In each case, the up spin and down spin band structures are degenerate. It's obvious that the tensile strain increase the band gap.

Fig.7 (a) Band gap of uniaxial strained ZGNR with different width (indicated by the number of zigzag chain in the transverse direction), inset is the normalized band gap versus strain curve, in which ZGNR with different width show similar linear dependence. (b) The spin polarization (up spin density minus down spin density) at the edges of ZGNR. Tensile (compressive) strain increase (decrease) spin polarization.

Fig.8 (a) The band gap of shear strained ZGNR with different width (indicated by the number of zigzag chain in the transverse direction). (b) The spin polarization (up spin density minus down spin density) at the edges of ZGNR. Shear strain always tend to decrease spin polarization at the edges.

Fig.A.1 The *k=0* Hamiltonian used in the Appendix. The left figure is the unstained Hamiltonian; the right figure is the strain induced perturbation Hamiltonian.

Fig.A.2 Band gap of strained *n=24* AGNR, comparison between numerical calculation (blue solid lines) and the analytical model (red dashed lines) developed in this section. In (a) edge distortion and 3rd nearest neighbor coupling are ignored, in (b) these two effects are included. It's well shown that the analytical model agrees well with the numerical simulation results. The deviation in higher strain region may be due to higher order effects.

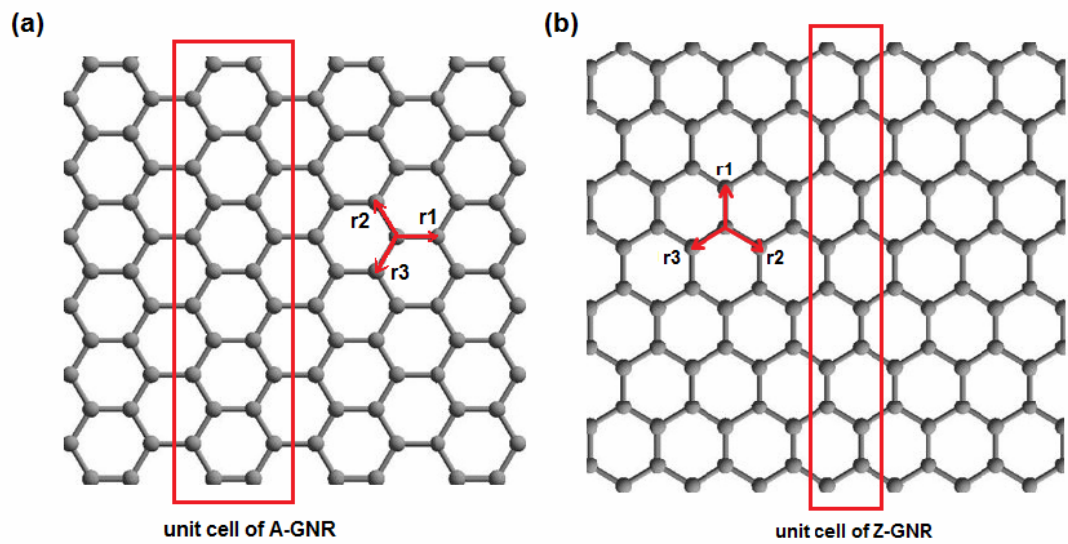

FIGURE 1

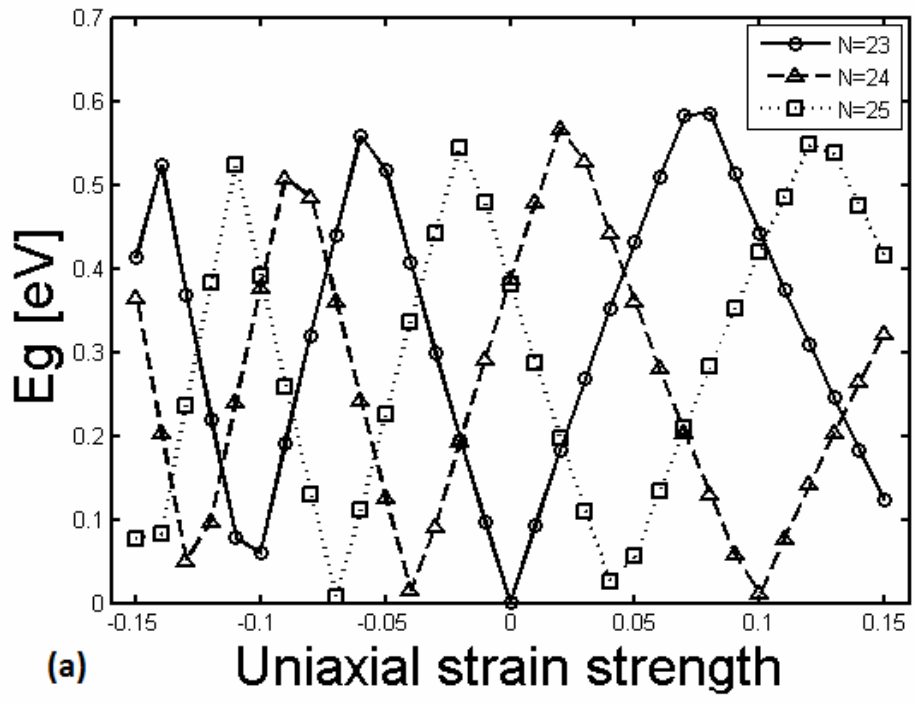
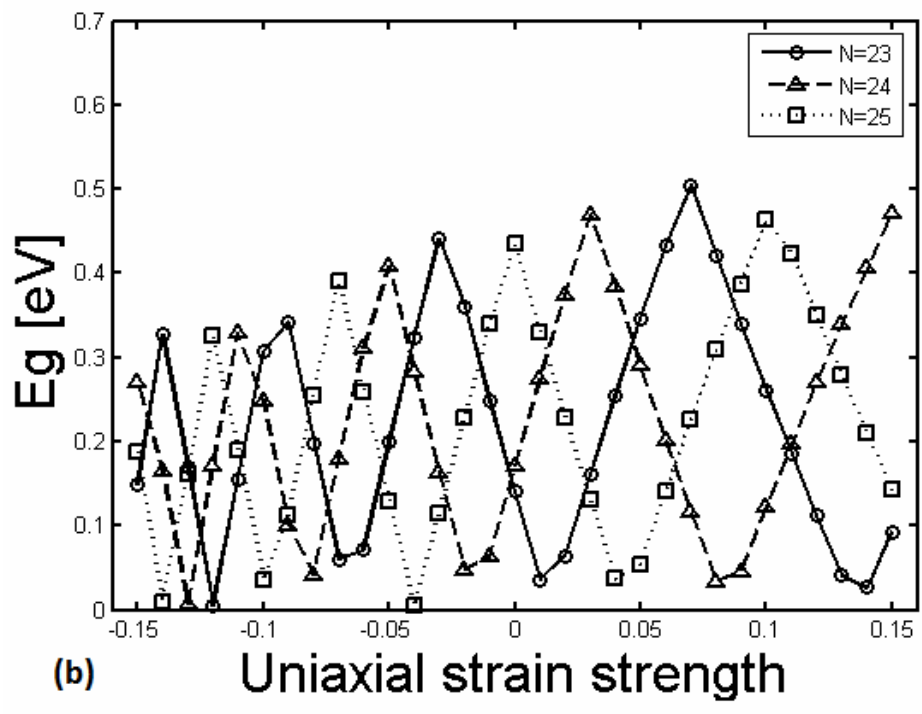

FIGURE 2

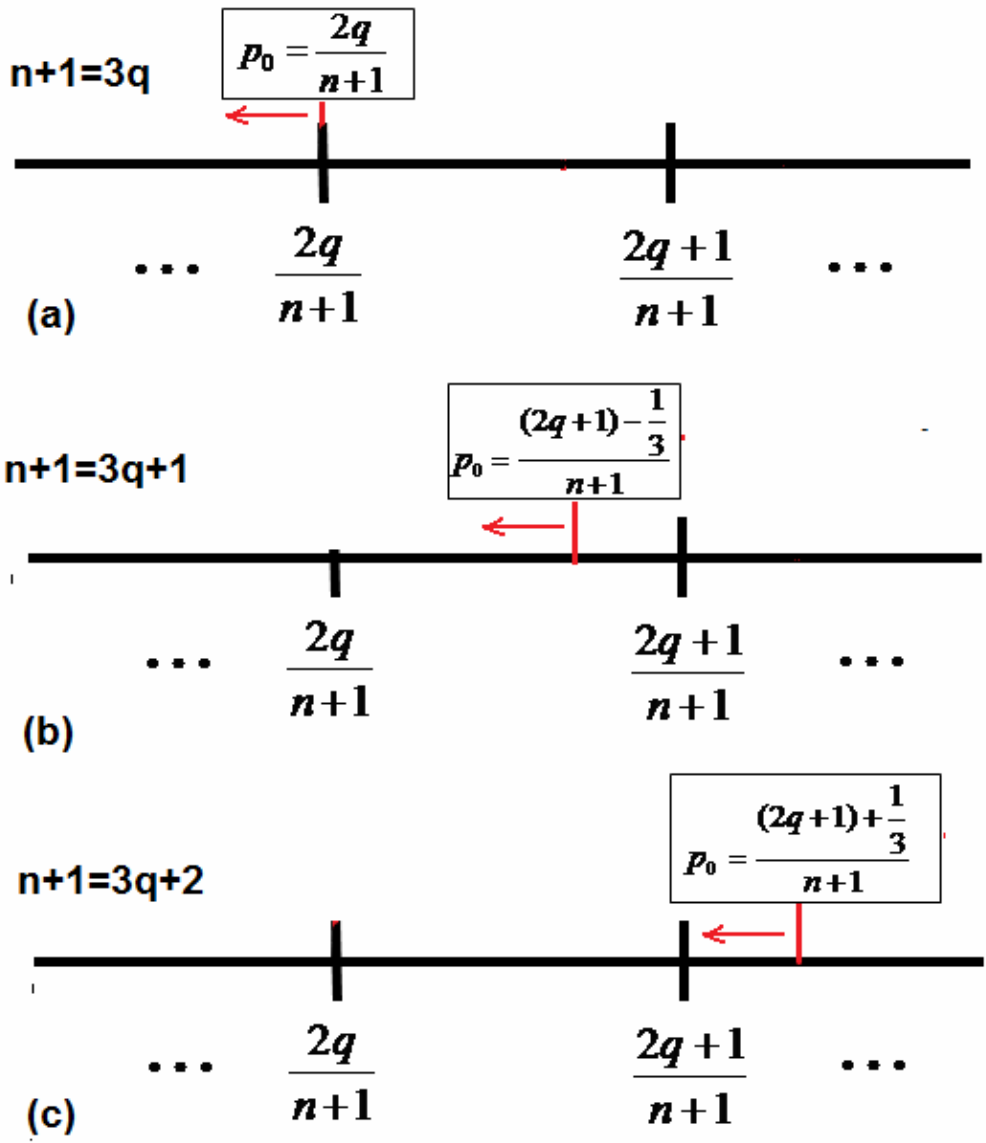

FIGURE 3

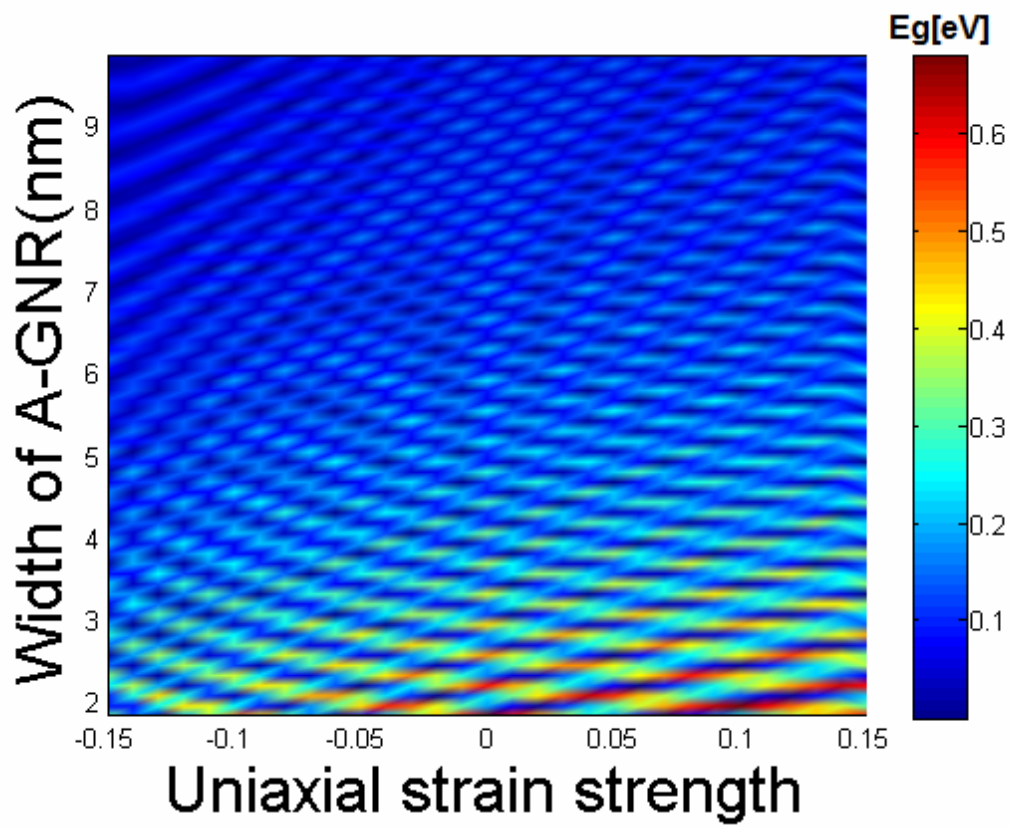

FIGUARE 4

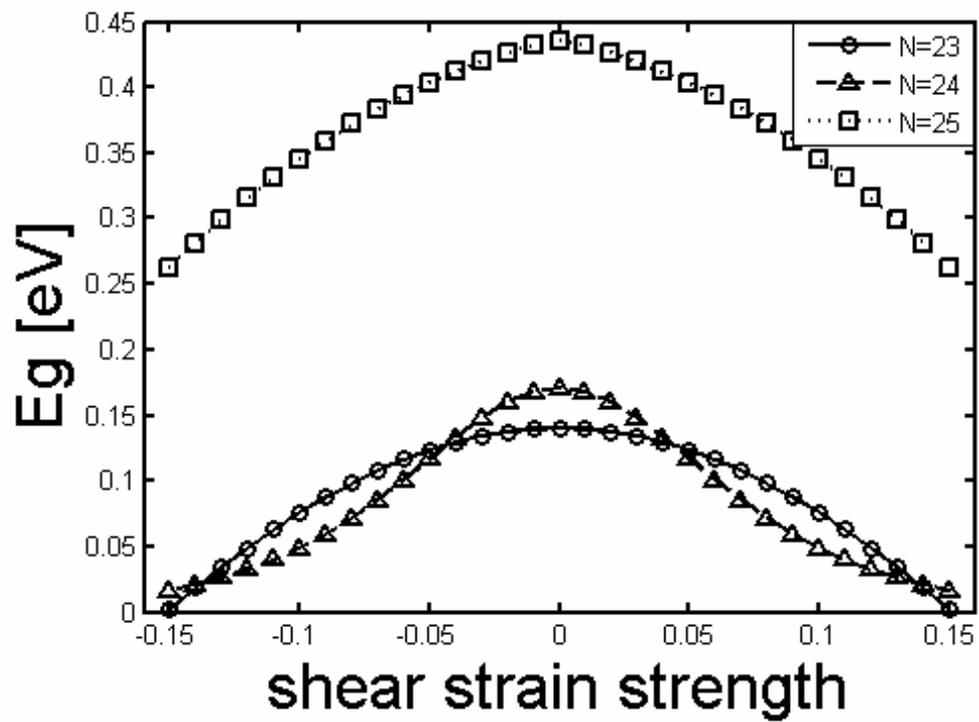

FIGUARE 5

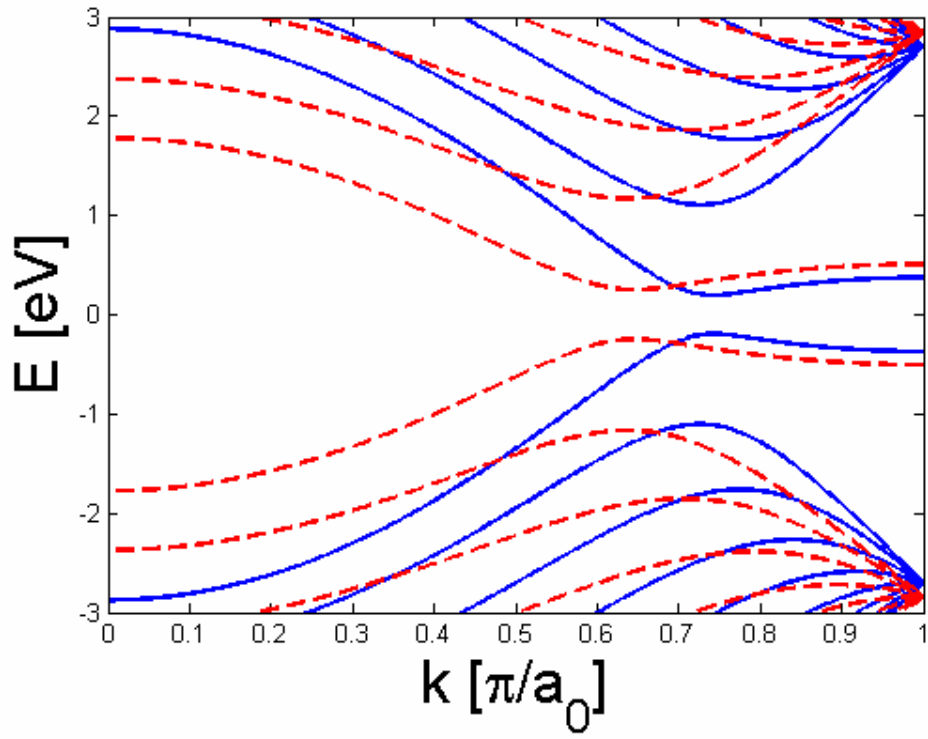

FIGUARE 6

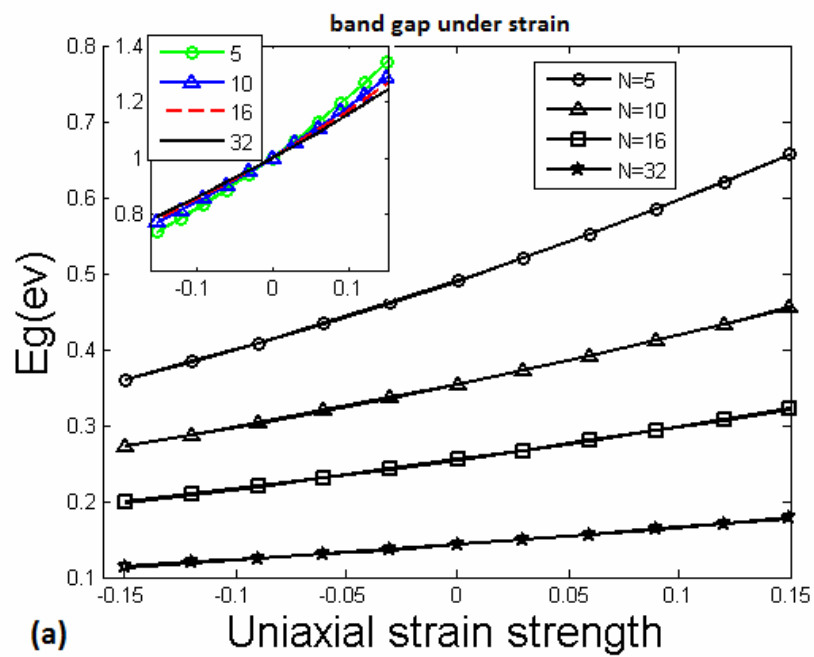

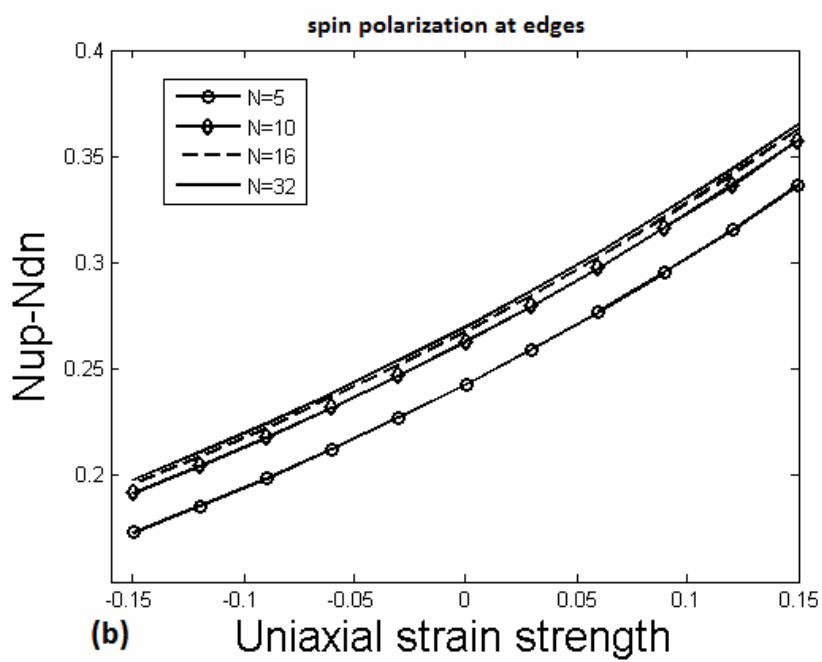

FIGURE 7

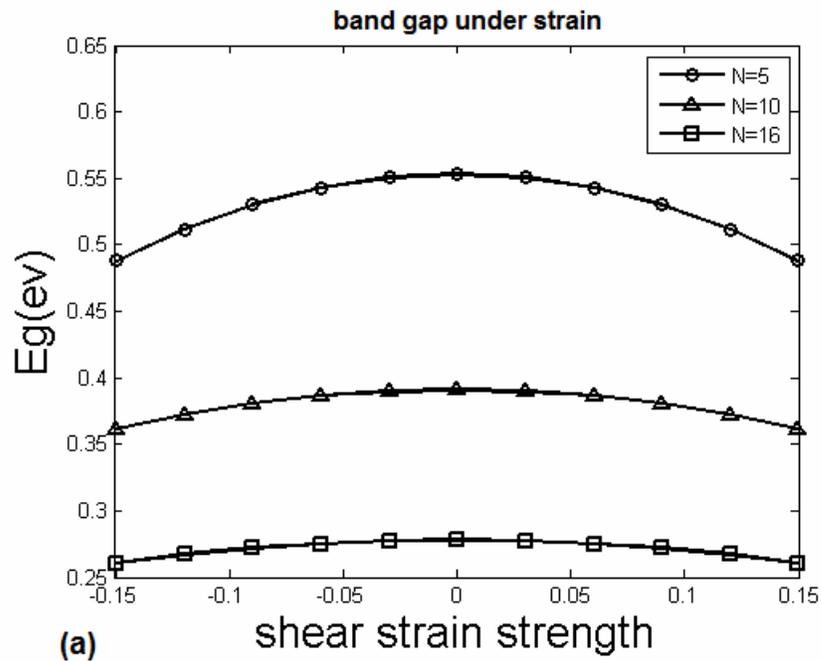

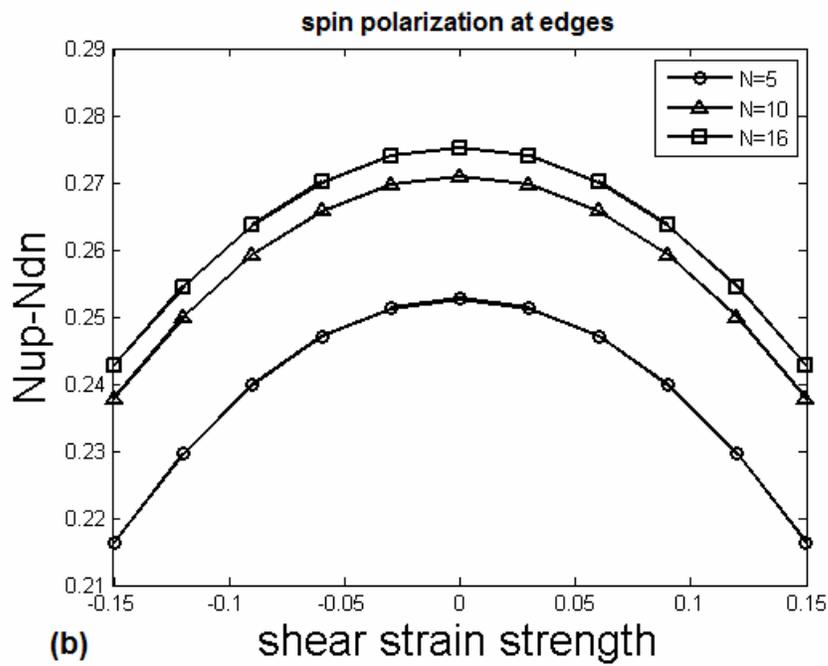

FIGURE 8

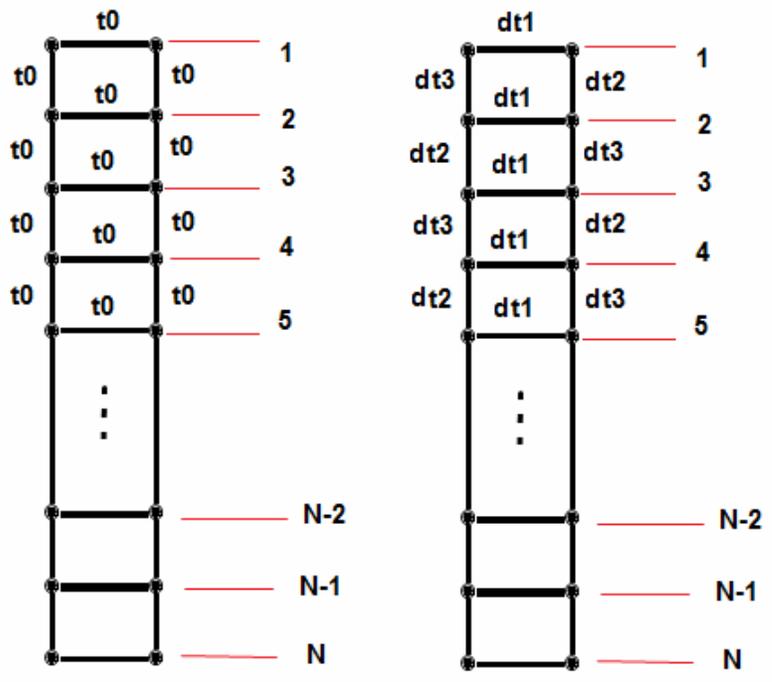

FIGURE A.1

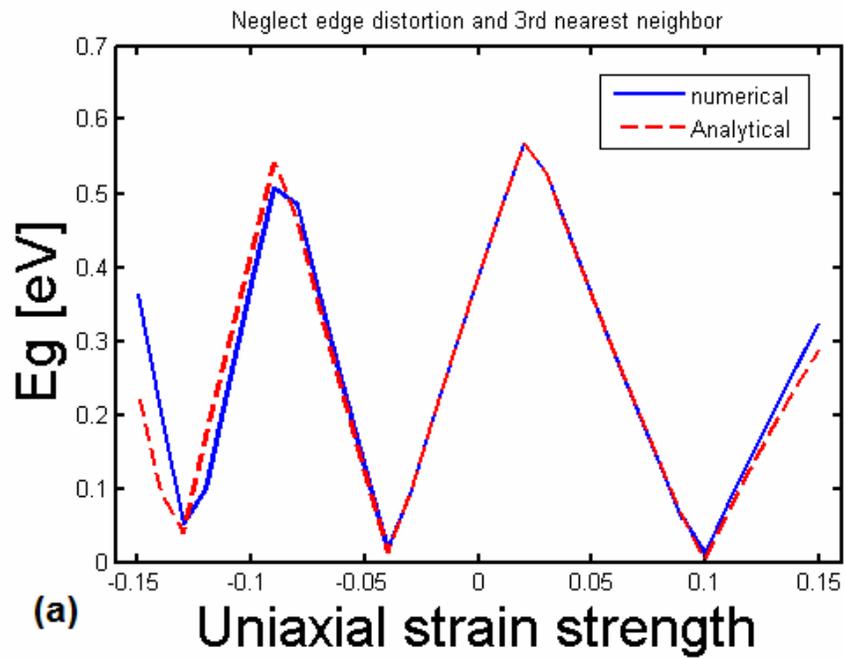

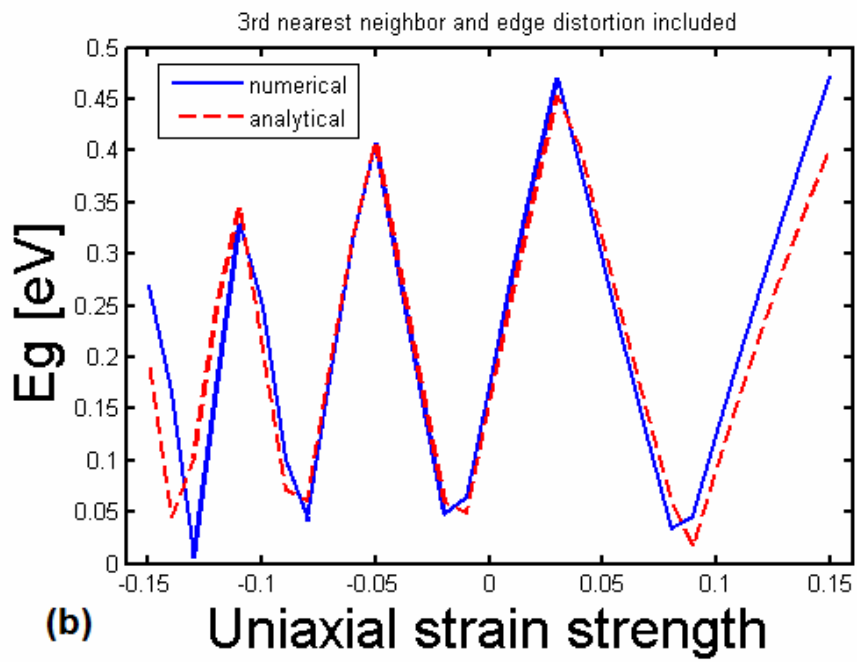

FIGURE A.2